# Uncited articles and their effect on the concentration of citations

Diego Kozlowski[1], Jens Peter Andersen[2] and Vincent Larivière[1,3]

[1] *diego.kozlowski@umontreal.ca*

École de bibliothéconomie et des sciences de l'information, Université de Montréal, Montréal, QC, Canada.

[2] *jpa@ps.au.dk*

Danish Centre for Studies in Research and Research Policy, Aarhus University, 8000 Aarhus C, Denmark.

[3] *vincent.lariviere@umontreal.ca*

École de bibliothéconomie et des sciences de l'information, Université de Montréal, Montréal, QC, Canada.

Observatoire des Sciences et des Technologies, Centre Interuniversitaire de Recherche sur la Science et la Technologie, Université du Québec à Montréal, Montréal, QC, Canada.

**Abstract**

Empirical evidence demonstrates that citations received by scholarly publications follow a pattern of preferential attachment, resulting in a power-law distribution. Such asymmetry has sparked significant debate regarding the use of citations for research evaluation. However, a consensus has yet to be established concerning the historical trends in citation concentration. Are citations becoming more concentrated in a small number of articles? Or have recent geopolitical and technical changes in science led to more decentralized distributions? This ongoing debate stems from a lack of technical clarity in measuring inequality. Given the variations in citation practices across disciplines and over time, it is crucial to account for multiple factors that can influence the findings. This article explores how reference-based and citation-based approaches, uncited articles, citation inflation, the expansion of bibliometric databases, disciplinary differences, and self-citations affect the evolution of citation concentration. Our results indicate a decreasing trend in citation concentration, primarily driven by a decline in uncited articles, which, in turn, can be attributed to the growing significance of Asia and Europe. On the whole, our findings clarify current debates on citation concentration and show that, contrary to a widely-held belief, citations are increasingly scattered.

**Introduction**

Several early bibliometric studies provided evidence that citations and publications follow power law distributions, implying high concentrations on one end and a long tail of low-yield observations on the other. Similarly, landmark studies by Lotka (1926) showed that a minority of researchers accounted for the majority of research publications. Bradford (1934) provided evidence that articles were concentrated within a core set of journals. Such concentration of research publications in a limited set of journals was key in the indexing practices of the Science Citation Index (Garfield, 1972). Subsequently, Seglen (1992) showed similar trends for citation distributions of journals, with only 15% of articles accounting for 50% of the citations.

Like Garfield's Journal Impact Factor, these early studies had the aim to help reference librarians maintain the most relevant and useful collections. This pragmatic view on maximizing the value of collections within budget restraints goes well together with the normative argument of Merton (1974) that scientists tend to reward those who came before them through their references, including the best and most relevant research. However, Merton (1968) also developed the concept of Matthew effect, wherein recognition is not only provided by scientific merit, but also influenced by researchers' symbolic capital, as well as by that of universities, journals, publishers, and scientific societies. This effect provides larger-than-expected citations to those articles which are already the most highly cited. Citations concentration is thus not only a practical way of assessing the usefulness of a library collection but more so an analysis of the degree to which hierarchies and sociological markers influence the scientific communication structures of a given field. This result has been crucial for the critical assessment of research evaluation based on citations metrics (Bornmann, Mutz, Neuhaus, & Daniel, 2008; DORA, 2012; Sugimoto & Larivière, 2018).

Early analyses of citation and reference concentrations focused on articles, countries, or journals, or on small samples of researchers. However, since the advent of large-scale, advanced author disambiguation algorithms (Caron & van Eck, 2014; D'Angelo & van Eck, 2020) it has become possible to also analyze citation concentrations at the author level at a large-scale. Nielsen & Andersen (2021) found that the most cited percentage of authors accounts for an increasing share of the total number of citations, even when factoring in the growth in number of publications and length of reference lists over time. However, the concentration of citations at the level of articles does not necessarily mirror the concentration of authors, mostly because of authors' different research productivity levels and evolving patterns of co-authorship.

While the high concentration of citations is widely acknowledged in the field of bibliometrics, there exists an ongoing debate regarding how citation concentration has evolved over time. This debate particularly revolves around the effect of electronic journals and large-scale bibliographic databases on the literature accessed and utilized by researchers. On the one hand, previous research has shown that researchers increasingly rely on older literature (Larivière, Archambault, & Gingras, 2008; Verstak et al., 2014) and on a diversity of journals (Acharya et al., 2014; Larivière, Lozano, & Gingras, 2014; Lozano, Larivière, & Gingras, 2012), which suggests that databases and digitalisation of research articles makes it easier for researchers to find and therefore cite a diversity of research articles, both old and new. Such improved access to a broader set of papers would therefore reduce the concentration of citations (Larivière, Gingras, & Archambault, 2009). Using reference lists from published articles and linear models, Evans (2008) provided opposite evidence, showing that researchers narrowed their citation practices as a consequence of digitization and increasingly rely on social markers of capital and preferential attachment, thus reinforcing citation concentration. Similar results were obtained by (Barabási, Song, & Wang, 2012).

The discussion of whether or not concentration is increasing over time is influenced by bibliometric aspects of the measurement of inequality, such as the increasing number of publications indexed in the databases (Kim, Adolph, West, & Stovel, 2020), and the inclusion of uncited articles (Pan, Petersen, Pammolli, & Fortunato, 2018; Wallace, Larivière, & Gingras, 2009). The contribution of uncited articles to the progress of science is a complex question; from a naïve point of view, an uncited article is of no use to the scientific community. However, we believe this is a simplistic view (Golosovsky & Larivière, 2021): such articles may provide inspiration, or they can serve as important background literature that is either forgotten, omitted due to space limitations, or questionable referencing practices. In the context of the Ortega y Gasset hypothesis (Ortega y Gasset, 1933), this perspective considers scientific advancements as building upon the collective efforts of a multitude of scientists. It emphasizes the incremental progress of science, which often relies on the numerous, sometimes overlooked or disregarded, contributions made by the majority of researchers. However, for concentration analyses, the omission of uncited articles is only logical under the assumption that uncitedness patterns are stable over time, or that uncited articles should not even be part of the scholarly record. Any change in the factors that affect uncited articles—such as the total number of citing articles—are likely to influence concentration. Frandsen & Nicolaisen (2023) recently showed that articles with few references are much less likely to be cited and that uncitedness varies between fields. Previous research has also shown that uncitedness changes over time (Pan et al., 2018;

Wallace et al., 2009), given the inclusion of new sources, increasing reference list lengths and the broader availability of scientific articles.

This article aims to contribute to the debate about the evolution of concentration of citations, emphasizing the role of uncited articles as a key factor in explaining variations in the concentration of citations. By showing the role of uncited articles in combination with reference-based and citation-based approaches, we shed new light on how certain methodological decisions can highly affect the outcome of analyses. We also consider citation windows, the changing size of bibliometric databases, and the role of self-citations to arrive at a robust explanation of how citation concentration evolves over time. Our results indicate that when uncited articles are included, there is a downward tendency in citation concentration. Given the relevance of uncited articles for the evolution of concentration, we also study the role of regions in the evolution of uncitedness and find that the increasing significance of Asia and Europe and their specific role among low-cited articles explain the reduction of uncitedness, and thus, inequality.

**Data & Methods**

The present analysis relies on publication metadata extracted from the WoS core collection, encompassing publications from 1980 to 2020. Two specific datasets were utilized from this source. The primary dataset, which forms the basis of the main analysis, focuses on articles published in journals that have been indexed throughout the entire period, as well as the references cited by these articles. This restriction to the core set of journals was implemented to mitigate the potential influence of newly incorporated journals on citation concentration. The second dataset encompasses all articles and references included in the complete WoS collection between 1980 and 2020, encompassing journals from various sources.

When conducting comparisons of received citations among different articles, it is crucial to address certain biases. Articles published in different years and fields cannot be directly compared due to inherent differences. For instance, older articles have had more time to accumulate citations, and articles in certain fields receive higher citation rates. To account for this, researchers commonly employ techniques such as time normalization, fixed citation windows and within-field normalization to ensure fair comparisons. These approaches help to mitigate the impact of publication year on citation counts and enable more accurate evaluations across articles. The number of publications covered by the Web of Science also grows over time, and so does the number of references (Andersen, 2023). Thus, for a given citation window, newer articles will have a higher chance of being cited. Fixed citation windows are thus also susceptible to bias which needs to be addressed with a year-normalization of citations. Finally, different disciplines have different citation patterns, and high variance in the degree of reference coverage in the database, with the best coverage in the natural and medical sciences and much less so in the social sciences and humanities (Andersen, 2023). Different fields have different dynamics in their citations: while in some fields the majority of citations are received in the first 3 years, in others it can take longer periods (Sugimoto & Larivière, 2018). These differences also need to be accounted for.

*Citation-based and reference-based approaches*

Two perspectives can be used for the compilation of concentration indicators: the citation-based and the reference-based approach. The citation-based approach is a forward-looking method that considers all articles published in a specific year and counts the citations they receive in subsequent years. It encompasses all articles, including those that did not receive any citations. This approach captures citations from different time periods. However, there is a well-documented increase in citations over time (Petersen, Pan, Pammolli, & Fortunato, 2019), which can distort annual comparisons of citation counts. For this reason, we control for citation inflation together with the widely used field- and year-normalization.

In contrast to the citation-based approach, the reference-based approach takes a backward-looking perspective. It considers all the references made by articles from a given year to articles published previously. This approach naturally excludes uncited articles from the analysis. However, it is also possible to include articles published within the defined time window that may not appear as references. This approach provides valuable insights into the referencing patterns and connections between articles across different publication years.

*Normalization*

We control all the elements mentioned above, following the normalization strategy proposed by Nielsen & Andersen (2021). For the citation-based method, we compute a year and field normalized number of citations, *nics*, and show all results for multiple citation windows. To compute *nics* we first calculate the weight of citations on year $y$, $\rho_y$ as:

$$\rho_y = \frac{1}{\sum ncits_y}$$

Where $ncits_y$ is the raw number of citations on year $y$. The weighted received citations of an article $i$ on the year $y$, $ics_{i,y}$ is

$$ics_{i,y} = ncits_{i,y} * \rho_y$$

and the total number of citations of the article $i$ in the dataset, $ics_i$ is

$$ics_i = \sum_y ics_{i,y}$$

the mean $ics_i$ for all $i$ in the discipline $k$, $mics_k$ is

$$mics_k = mean(ics_k)$$

Finally, the normalized citations $nics_i$ for the article $i$ that belongs to the discipline $k$ is

$$nics_i = \frac{ics_i}{mics_k}$$

Unless mentioned otherwise, we used $nics_i$ citations for all the following analyses.

In the reference-based approach, the normalization by year is not required since all references under consideration are made during the same year. Consequently, concerns regarding citation inflation can be alleviated as the distribution of references remain unaffected by the passage of time. Nevertheless, different disciplines have different referencing patterns, so normalization by field is necessary.

Each article produces a certain number of references, *nref*. The average number of references in articles of a field *k* is:

$$mref_k = mean(nref^k)$$

We use this as the normalization factor. Articles from previous years receive $ncits_k^y$ on year *y* from discipline *k*. To weigh the citations, we divide the citations from each field by the average of that field, before summing citations from different fields.

$$nref_i^{norm} = \sum_k \frac{nref^k}{mref_k}$$

As all citations come from the same year, and we use time windows to control the time distance between articles and references, we do not normalize by year.

*Citation windows*

To compare articles within the same time range, we employ citation windows ranging from 2 to 10 years, excluding the year of publication. This exclusion is necessary to mitigate potential discrepancies arising from variations in the month of publication. For instance, articles published in December may have a different citation count compared to those published in January when considering a one-year citation window. The decision to exclude articles from the same year enhances comparability but also results in the exclusion of citations. Consequently, the proportion of uncited articles may increase, particularly for those that were solely cited during their year of publication.

In the citation-based approach, we establish a year limit and exclude articles that are younger than the defined citation window. To ensure consistency and preserve the integrity of the time window, this exclusion applies to articles published after a specific threshold. For instance, when using a citation window of ten years and considering a dataset that spans until 2020, articles published after 2011 are excluded from the analysis.

On the other hand, the reference-based approach employs a backward-looking time window. For instance, a one-year time window includes articles published in the year preceding the reference. Conversely, to maintain consistency, articles that predate the citation window are excluded from the analysis. Beginning from the inception of our dataset in 1980, this means that for a ten-year citation window, articles published in 1989 or earlier are excluded from the analysis.

*Inequality measure*

Citation inequality is measured using the Gini coefficient. This choice is driven by the fact that the Gini coefficient captures the concentration of citations across all articles, providing a comprehensive view of inequality. In contrast, alternative measures such as the proportion of citations accumulated by the top x% focus solely on the concentration within highly cited articles, neglecting the relative concentration among articles with lower to moderate citation counts. However, we do acknowledge the importance of these alternative metrics and employ them as a robustness check to validate our analysis. By incorporating multiple measures, we ensure the robustness and reliability of our findings concerning citation concentration and inequality.

*Uncited articles*

The citation-based and reference-based approaches inherently differ in their treatment of uncited articles. In the citation-based approach, uncited articles are naturally included as the data retrieval process begins with all articles published in a given year. Conversely, the reference-based approach naturally excludes uncited articles as the analysis focuses on the references made in a given year. Furthermore, it is important to note that the number of uncited articles has exhibited a downward trend over time. For instance, when considering a ten-year citation window, the proportion of uncited articles decreased from 34% in 1980 to eleven percent in 2010 across all articles in the WoS database, excluding self-citations. Recognizing the potential impact of uncited articles, we conduct analyses using both the citation-based and reference-based approaches with and without the inclusion of uncited articles. This allows us to compare the outcomes by either incorporating or filtering out uncited articles, thus capturing the potential effects they may have on the results.

**Results**

Figure 1 presents the results obtained for the combinations of citation- and reference-based approaches, including and excluding uncited articles, and limited to the core set of journals. We consider multiple time windows for each method, although these are forward-looking for citation-based concentration and backward-looking for reference-based concentration (see SI Figure S1). The difference between the citation-based and reference-based approach is significant, yielding completely different perspectives on the evolution of the concentration of citations.

Figure 1a shows the Gini index for each approach on its own scale, with a different color scheme for each time window. The citation-based approach, including and excluding uncited articles, always shows a decreasing trend, which suggests a decrease in concentration. We can also see that the range of the Gini is smaller when we exclude uncited articles (0.45-0.60) than when they are included (0.60-0.70). For the reference-based approach, the tendency completely changes with the inclusion of uncited articles. Time windows are hardly comparable on the reference-based approach, as in any given year they are pointing to a different subset of articles, while the citation-based approach always shows the same articles, with a larger or smaller time span for gathering citations. When uncited articles are included, the reference-based approach shows a behavior similar to citation-based concentration and very large inequalities (the Gini moves between 0.73 and 0.8). Figure 1b shows the same as Figure 1a, but on a common scale and with color schemes referring to the analytic approach.

Overall, those results show that the approach chosen influences I) the effect of citation windows (Figure 1a), II) the level of inequality (Figure 1b), and III) the conclusion regarding historical trends in concentration (Figure 1c). We expect a higher Gini index when including uncited articles, as the cumulative share of citations for those articles is 0%, enlarging the lower bound of the Lorenz curve that builds the Gini index (see Figure S2). The reference-based approach gives the lowest values of inequality when excluding uncited articles, and the highest when including them. Excluding uncited articles also reduces the Gini for the citation-based approach, but the difference is not as large.

Figure 1c shows the difference in Gini index from the first to the last year of analysis for each time window. When including uncited references, both the citation-based and reference based-approach converge to show that, between 1980 and 2020, there was a reduction in the concentration of citations between 5% and 8%, depending on the time window used. When we exclude uncited articles, the citation-based approach finds a reduction or stagnation in the concentration for almost all time windows, between 0% and 2.5%. For the reference-based approach, the omission of uncited articles reverts the conclusion and shows an increase in the concentration between 4% and 7%. The rise is especially large for the shorter time windows.

For smaller values, the distribution of citations shows a cardinality problem. As shown by the Lorenz curves (Figure S2), the part of the distribution that considers cited articles is smooth but abruptly cut on the zero-lower bound. The probability of an article receiving at least one citation is dichotomous and heavily influenced by the total number of articles that generate references (Wallace et al., 2009). On the other hand, the distribution among cited articles is smoother. The growth in the database increases the probability of articles receiving at least one citation, moving a growing proportion of articles from the uncited to the cited set. If uncited articles are not included, the evolution over time will show a set of articles that disproportionately grows from the lower bound, while the increases in the highly cited article only grow smoothly. This is why the reference-based approach excluding uncited articles shows an increase in inequality.

On the whole, results obtained with the four approaches can be divided into three time periods with distinct behavior: from 1980 until 1990, between 1990 and 2005/2010, and from 2005/2010 until the end of the series. The 20 years in the middle period show a decrease in concentration of citations, while the others either show a stagnation or an increase, depending on the method. These four distinct approaches, although capturing the same underlying phenomena, exhibit varying sensitivities to the fluctuations of each period. As a result, they yield different conclusions regarding whether the concentration of citations is increasing or decreasing over time. It is noteworthy that the downward trend in concentration is consistent across different academic disciplines, as demonstrated in Figure S3.

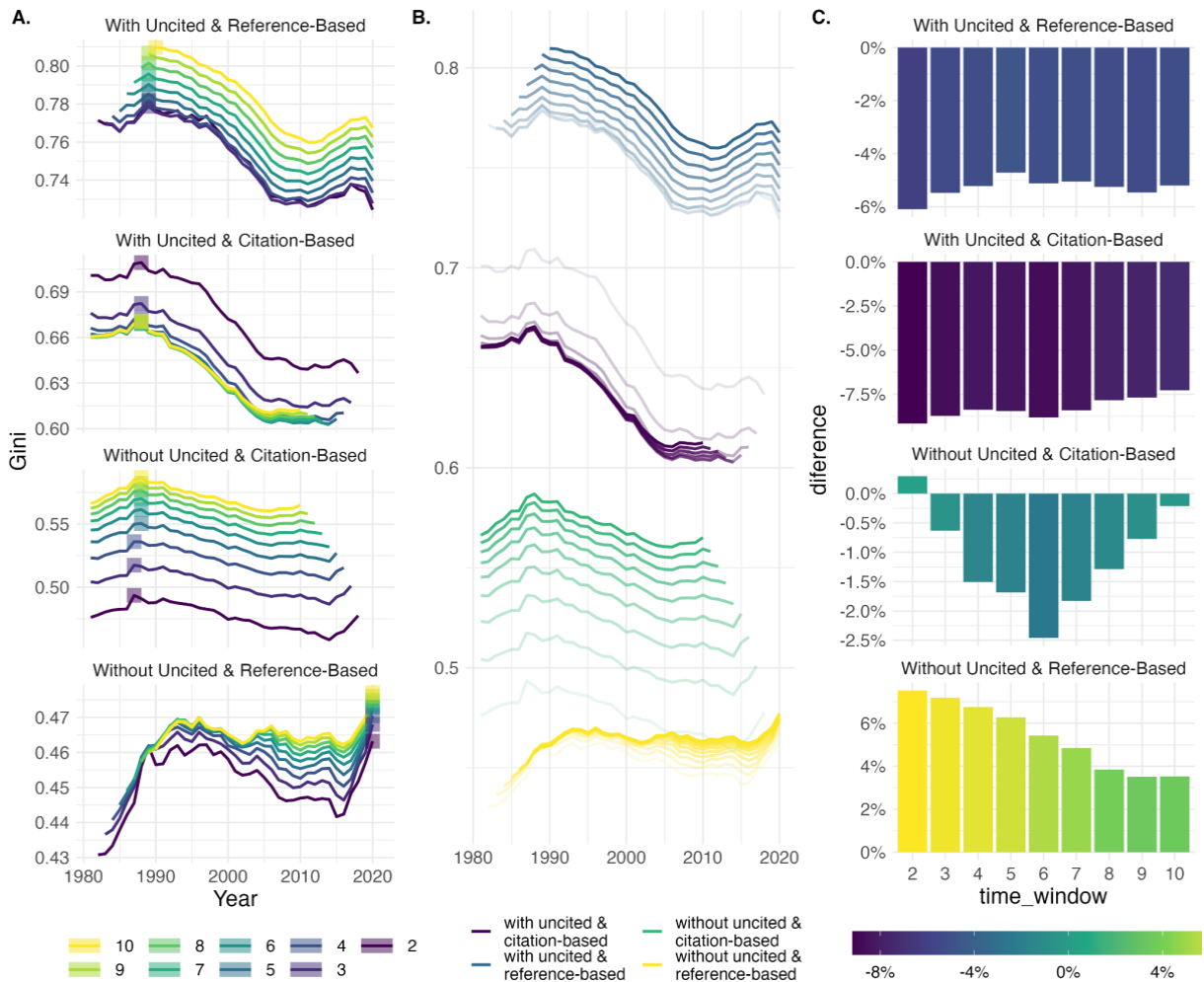

Figure 1. Evolution of Gini by time window and publication year. Citation-based and reference-based approaches, with and without including uncited articles, for the core set of journals. Each method by time window and peak highlights (A), different methods on a common scale (B), and the end-to-end total change by method and time window (C)

The presence of uncited articles plays a significant role in explaining the changes observed in citation inequalities over time. In order to gain a comprehensive understanding of the impact of self-citations and the influence of uncited articles on citation patterns across the entire Web of Science (WoS) database, Figure 2 is provided. This figure presents the evolution of uncited articles for both the complete WoS dataset and the core set of journals, as well as the proportion of uncited articles with and without considering self-citations. When self-citations are excluded, there is a substantial increase in the number of uncited articles observed for both the entire WoS (2A vs 2C) and the core set of journals (2B vs 2D). It is important to note that the core journal set generally exhibits a lower proportion of uncited articles, whether self-citations are included (2A vs 2B) or excluded (2C vs 2D). Nevertheless, in all cases, there is a notable decline in the share of uncited articles over time. Based on these findings, it can be inferred that the Gini inequality is expected to continue its downward trend across the entire WoS database, even when self-citations are removed from the analysis. This observation highlights the significance of uncited articles in shaping citation inequalities.

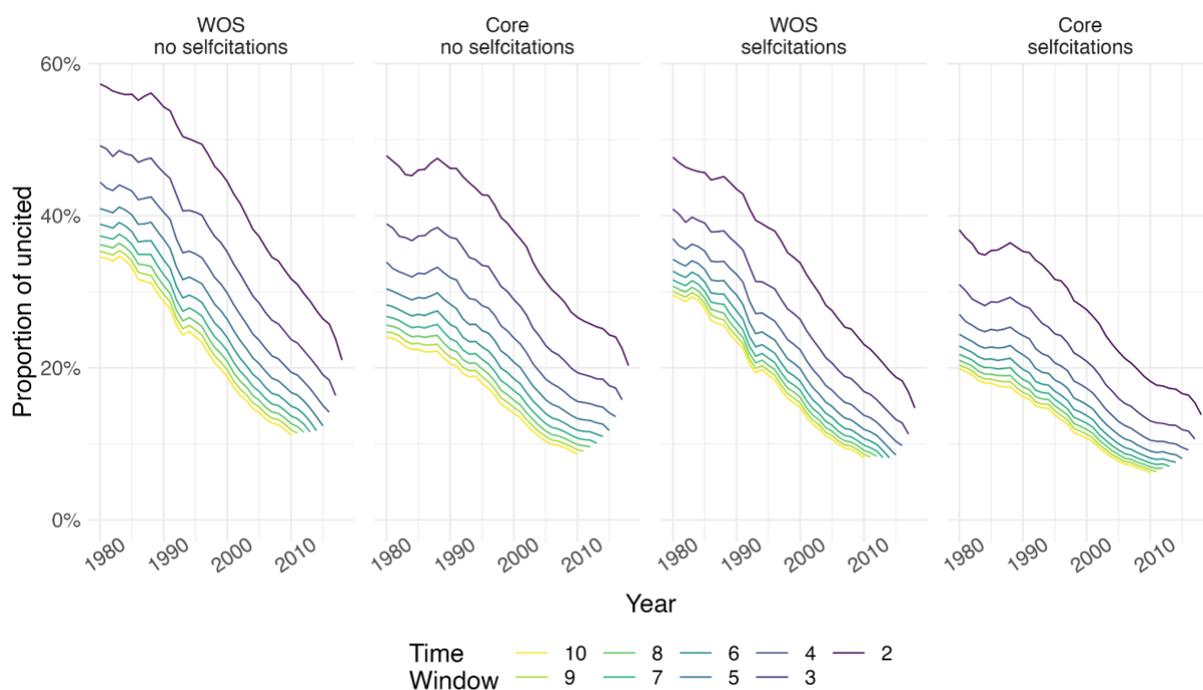

**Figure 2. Proportion of uncited articles by time window. All WoS, without self-citations (A), core set of journals, without self-citations (B), all WoS, including self-citations (C), core set of journals, including self-citations (D)**

The increased coverage of WoS can also influence uncitedness and, therefore, the concentration of citations. As seen in Figure 2, when we restrict to the journals that were covered since the beginning of the period, we still see a decrease in uncited articles. Self-citations are another potential explanation but were also discarded in Figure 2. Figure 3 explores the role of authors' regions on uncitedness. The figure shows what would happen with the proportion of uncited articles if all articles and references from a region were removed from the dataset.

Removing a region's articles has two opposite effects: 1) reducing all references made by articles from that region, increasing the probability of the remaining articles becoming uncited, as well as 2) reducing the number of uncited articles from that region. Figure 3 shows that the first effect is largely dominant, except for Africa where both effects are roughly even. If a region provides many citations to lowly cited articles from other regions, and its proportion of uncited articles is not large, then removing that region from the analysis creates an increase in uncitedness. The size of the effect will be dependent on the proportion of paper that comes from this region. Figure 3 shows the important effect that North America had at the beginning of the time series, but also the decrease in importance during the 90's given by the increasing contribution of articles from Europe and Asia.

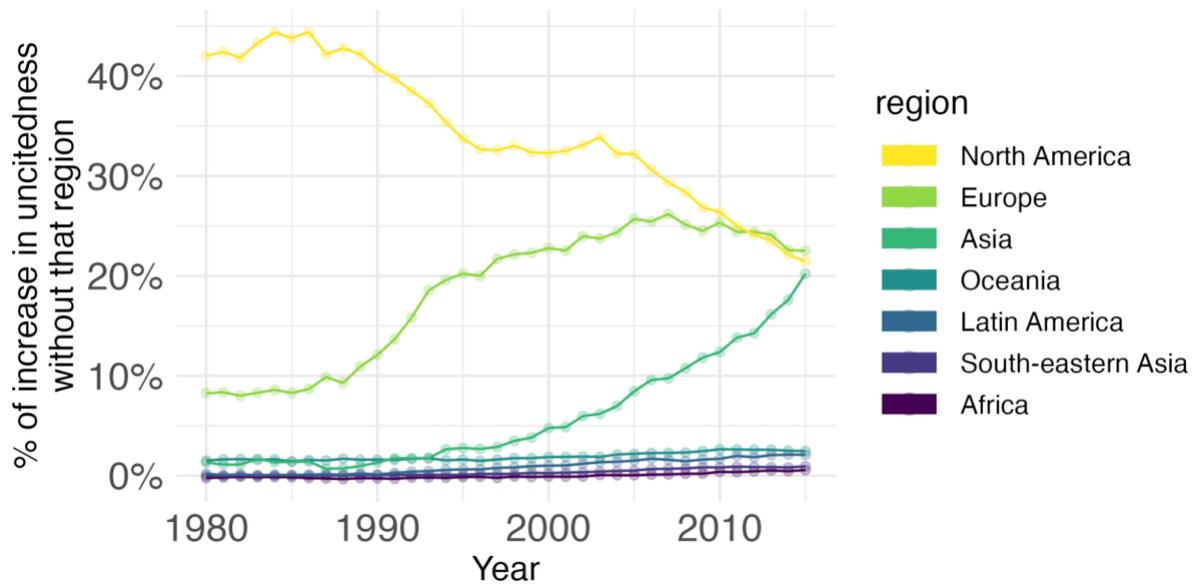

**Figure 3. Percentage change in the uncited articles when articles and references from a region are removed from the database. Five-year time-window.**

Such an effect is mainly driven by the number of articles from each region and the increased percentage of non-US articles in WoS. But it also implies a different pattern in the distribution of references and citations generated by each of these three regions, which all combined create a steep decrease in the percentage of uncited articles in the core WoS dataset. Figure 4 shows how each region cites and gets cited in the subset of highly cited articles (the top 1% most cited) and the subset of single-cited articles. The role of North America in low-cited articles is not as relevant as in top-cited articles. This is true for the proportion of low-cited articles that come from North America, and the proportion of references made by North American articles that account for the single citation of an article. The latter indicator is the main driver of changes in the proportion of uncitedness, and as we explained above, in the changes on the Gini index. Both Europe and Asia have a more central role in the reduction of uncitedness and, therefore, the increased relevance of those regions in the WoS database has an effect of reducing citation concentration. For top-cited articles, the role of North America becomes more important. This region accounts for the largest share of top-cited articles, although Asia shows a high increase in the series' final years. Nevertheless, when we analyze where the citations from top-cited articles come from, we can see that Europe and Asia account for a larger share of the distribution.

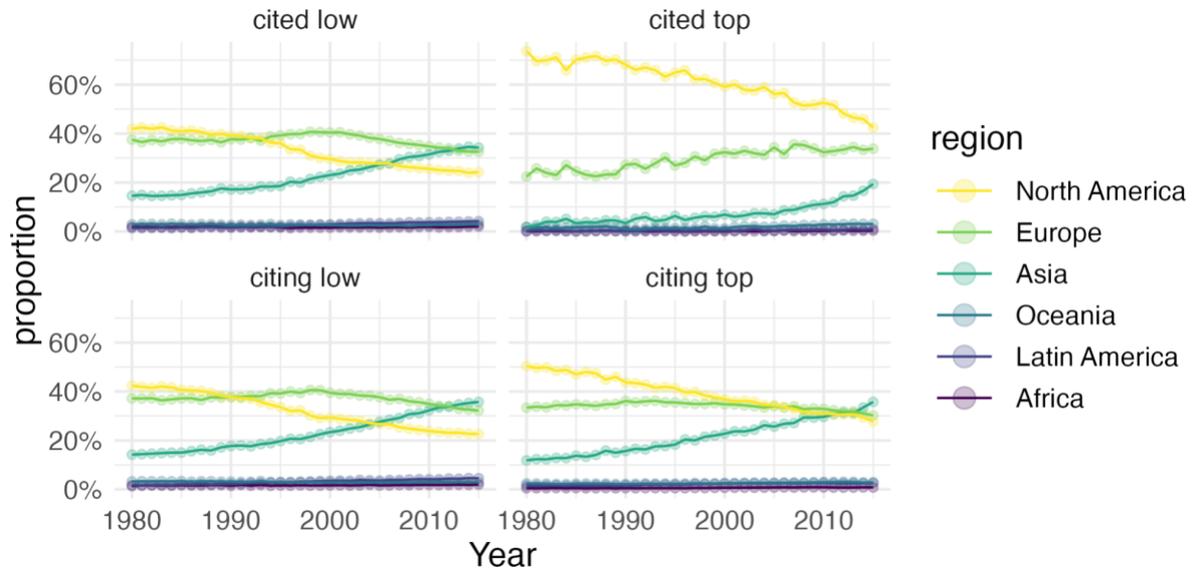

**Figure 4. Evolution of citations by region in the most and least cited articles. Cited low: proportion of articles with a single citation. Cited top: proportion of articles on the top 1% most cited. Citing low: proportion of the articles that cite a single-cited article. Citig top: proportion of the articles that cite a top 1% most cited article.**

To complement the Gini index, we also examine how the concentration of citations evolves for top-cited articles. Following the citation-based approach, Figure 5 shows the proportion of citations accumulated in the top 1%, 5%, and 10% most cited articles each year. Here, the denominator is the total number of citations for articles published each year, so uncited articles do not play a role. We can nevertheless see a decreasing trend analogous to that observed in Figure 1 for the citation-based approach, excluding uncited articles. The top 1% most cited articles show a stable trend around 10% of accumulated citations, while the top 5% and 10% most cited show a decreasing trend, from 34% to 30%, and 50% to 44% respectively, on the two-year citation window.

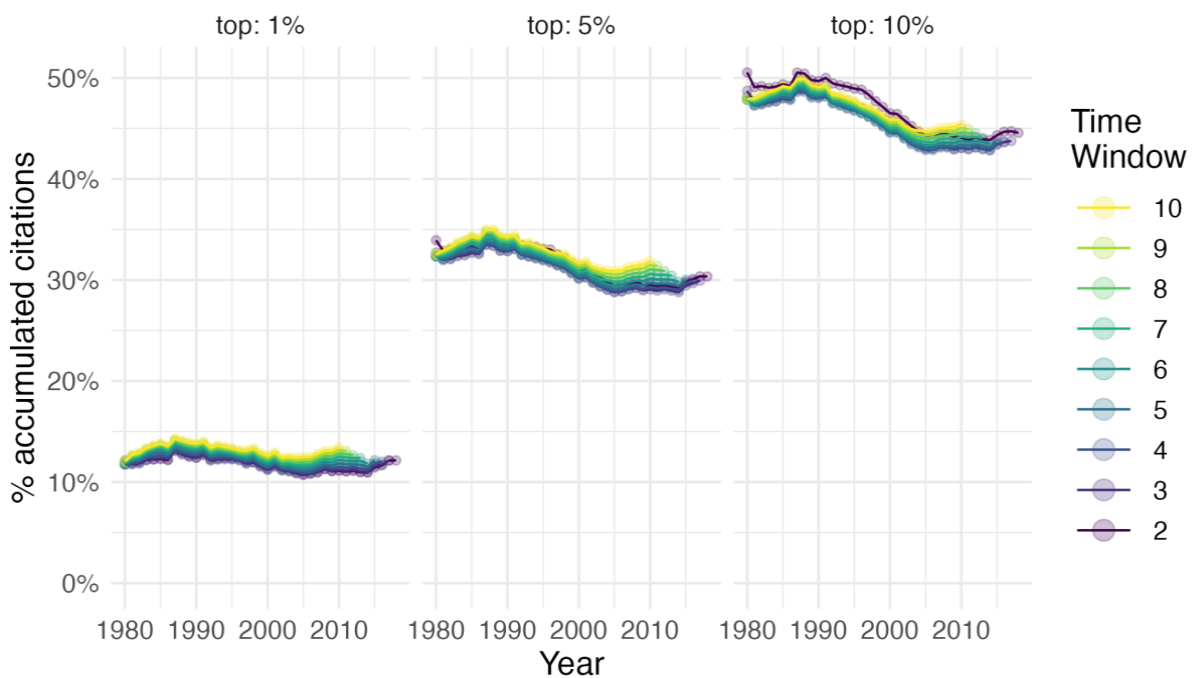

**Figure 5. The proportion of accumulated citations in the top 1%, 5%, and 10% most cited articles, over time, by time window.**

**Discussion and Conclusion**

Concentration of citations is a persistent issue in science. The Gini index shows levels of concentration that are much higher than those of income inequality, which are widely seen as very problematic (Chancel & Piketty, 2021). This research article highlights the importance of the methodology employed in measuring the concentration of citations and its implications for drawing conclusions about historical trends, with a particular focus on the significant influence of uncited articles. The citation-based approach, by considering all articles published in a given year, inherently incorporates uncited articles. On the other hand, the reference-based approach, which examines articles referenced in a given year, does not naturally account for uncited articles. Our results show that, when including uncited articles, we observe a decrease in concentration. Conversely, when uncited articles are excluded, an increase in concentration is observed.

The findings of this article emphasize the significance of considering uncited articles and their potential influence on citation concentration. It underscores the need for greater clarity and transparency in addressing the omission of uncited articles within the reference-based approach. By clarifying the impact of uncited articles on concentration dynamics, this research contributes to a more comprehensive understanding of citation patterns and interpretation of studies that suggest that citations are increasingly concentrating (Barabási et al., 2012; Evans, 2008), and explains why those including uncited papers showed that concentration were decreasing (Larivière et al., 2009). Our analysis showed that the most robust method for the analysis of citation concentration over time is the citation-based approach, considering the uncited articles, with a field and year normalization, and a fixed time window. The expansion of the bibliometrics databases and the evolution of self-citations needs also to be addressed. The results of this method show a consistent decrease in the concentration of citations.

Although it could be argued that the inclusion of uncited articles or even low-cited articles is a methodological choice, we consider their inclusion necessary. Those uncited papers were accepted by journals—which believed that they had something to contribute to science—and are part of the scholarly record. Moreover, the omission of uncited articles, along with the expansion of the WoS and the inflation in the number of citations artificially affects the trends in concentration by focusing on one part of the citation distribution. While the decrease in uncited articles may be influenced by the digitalisation of journals, better indexation and the increased coverage of bibliometric databases, we have also found that different regions play different roles in their contribution of uncitedness. The decreasing relative citation rates of North American papers and the increased participation of Europe and Asia affects citation patterns. While North America still contributes to the greater part of the top 1% most cited articles and produces many of the citations those articles receive, its role among articles cited only once is much lower. Therefore, by including uncited articles we can arrive at a more robust conclusion over the global trend, and the study of single-cited articles help understand why the trend changes over time.

# Supplementary material

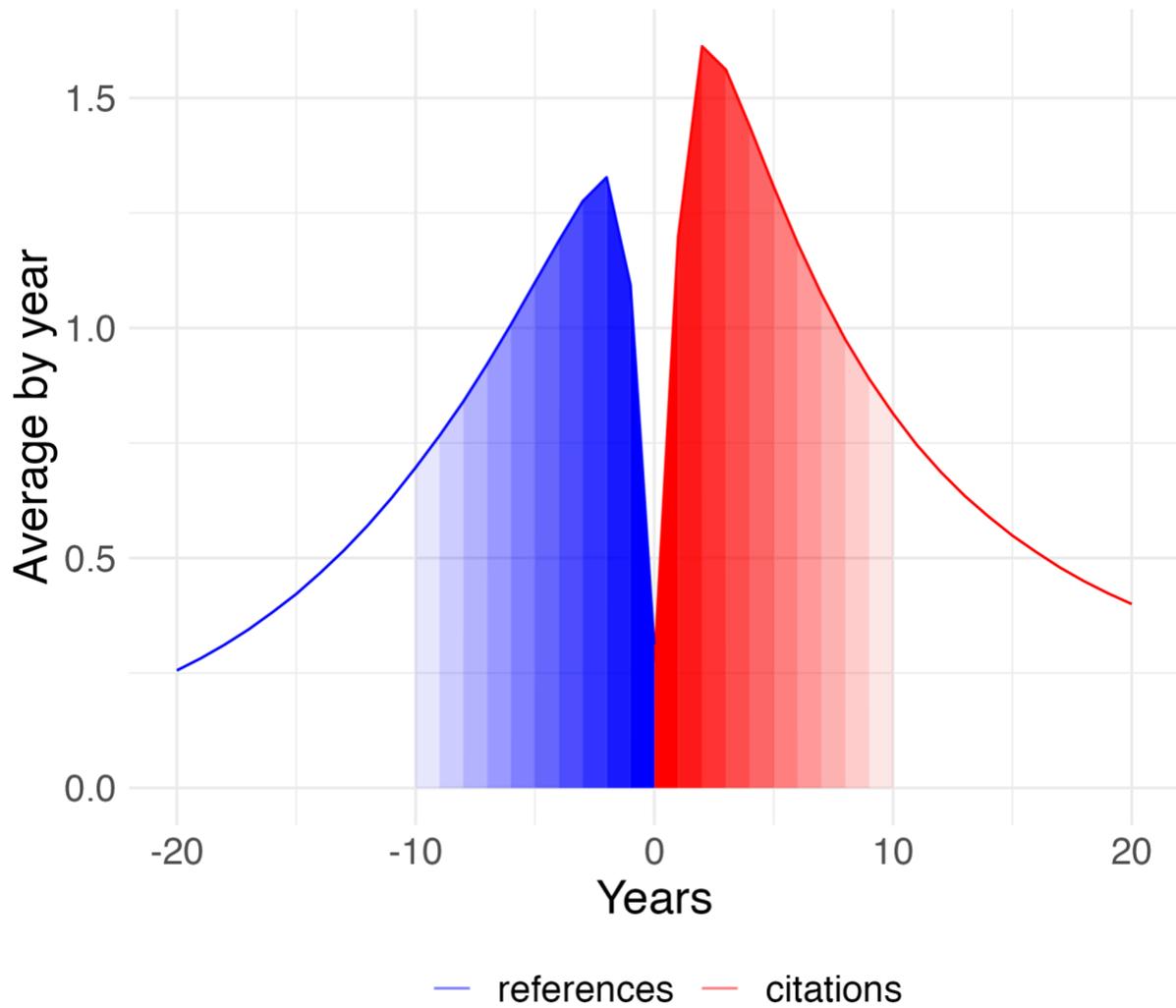

**Figure S1. Average citations (references) by time span after (before) publication.**

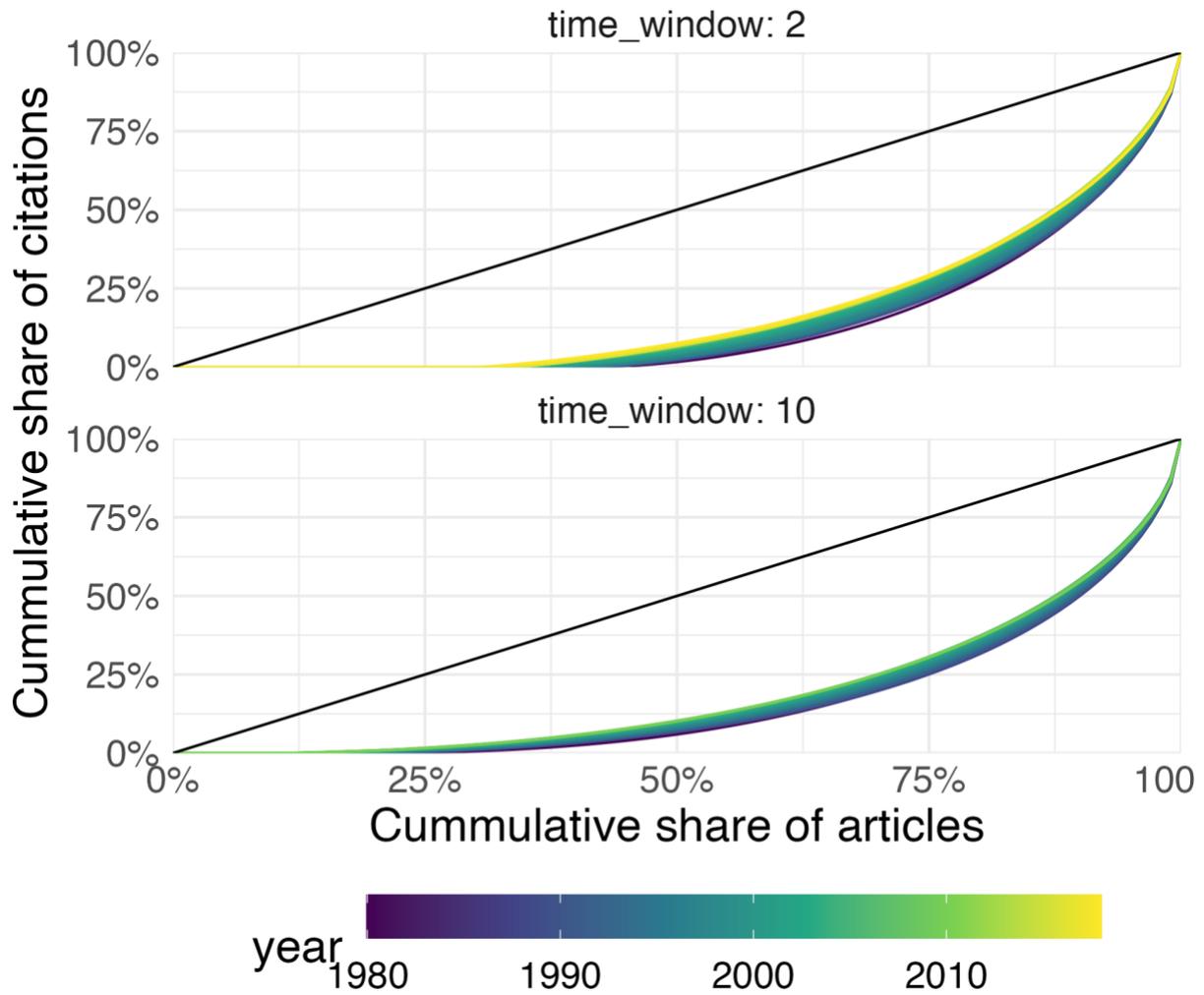

**Figure S2. Lorenz curves by year. For two and ten years' time-windows**

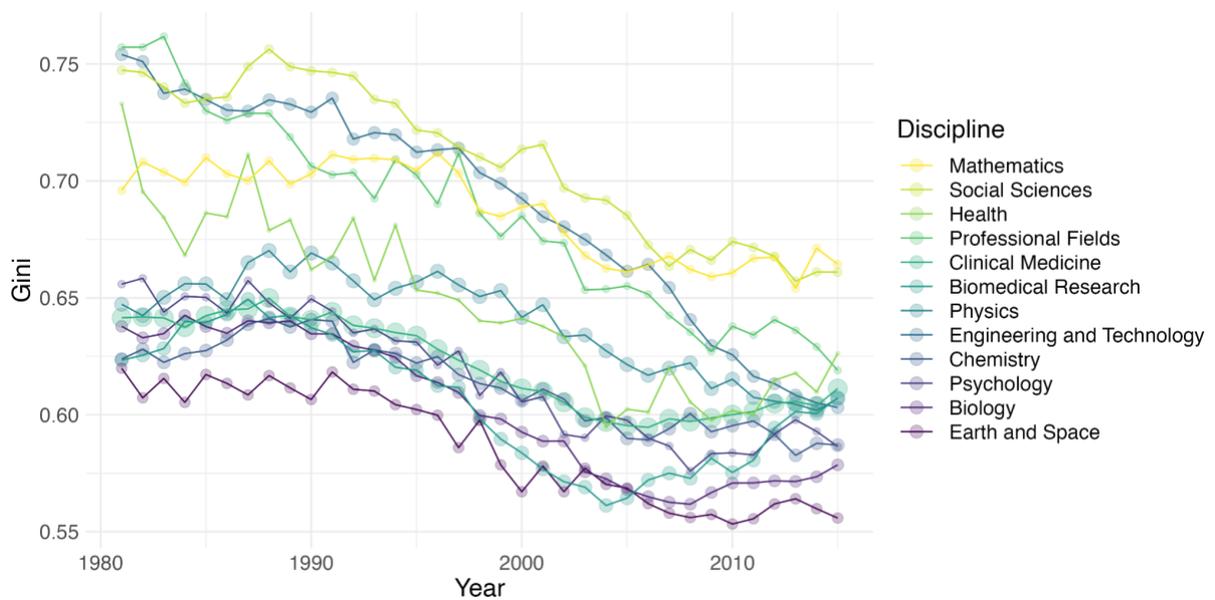

**Figure S3. Evolution of Gini by discipline for a five years time-window, using a citation-based method.**